\documentclass[preprint]{aastex}

\usepackage{graphicx}

\newcommand{\be}{\begin{equation}}
\newcommand{\ee}{\end{equation}}
\newcommand{\nn}{\mbox{} \nonumber \\ \mbox{} }
\newcommand{\ba}{\begin{eqnarray}}
\newcommand{\ea}{\end{eqnarray}}

\newcommand{\Alfven}{ Alfv\'{e}n }

\newcommand{\Bf}{{magnetic field\,}}

\newcommand\etal{\textit{et al.\ }}
\newcommand\eg{\textit{e.g.\ }}
\newcommand\cf{\textit{cf.\ }}

\newcommand\lo{\mathrel{\raise.3ex\hbox{$<$}\mkern-14mu\lower0.6ex\hbox{$\sim$}}}
\newcommand\go{\mathrel{\raise.3ex\hbox{$>$}\mkern-14mu\lower0.6ex\hbox{$\sim$}}}

\begin{document}
\date{}  
\title{On the nature of eclipses in binary pulsar J0737-3039}
\author{
M.  Lyutikov}
\affil{Department of Physics, McGill University,
Montr\'eal, QC, Canada\\ and \\
Canadian Institute for Theoretical Astrophysics,
University of Toronto, Toronto, ON, Canada}

\begin{abstract}
We consider magnetohydrodynamical  interaction
between relativistic
 pulsar wind and static magnetosphere in  binary pulsar system PSR J0737-3039.
We  construct  semi-analytical model  describing the  form of the interface separating
the two pulsars.
 An assumption of  vacuum dipole  spin down
for Pulsar B leads to eclipse duration ten times longer than observed.
We discuss possible Pulsar B  torque modification and magnetic field estimates
 due to the interaction with
 Pulsar A wind.
 Unless the orbital inclination is $\leq  86 ^\circ$, the 
duration of eclipses is typically shorter than the one implied by 
the size of the eclipsing region. 
We  propose  that  eclipses occur due to  synchrotron absorption by mildly
relativistic particles in the shocked Pulsar A wind.  The corresponding
optical depth may be high enough if  Pulsar A wind density is at the upper allowed limits. 
We derive jump conditions at oblique,  relativistic, magnetohydrodynamical shocks
and discuss the  structure of 
 the shocked  Pulsar A wind. Finally, we speculate on a  possible mechanism
of orbital modulation of Pulsar B radio emission.
\end{abstract}
\keywords{pulsars: individual (PSR J0737-3039) -- stars: neutron  -- shock waves}

\section{Introduction}

A recent discovery of  eclipsing binary pulsar system PSR J0737-3039 \citep{burgay03,lyne}
may serve as an important tool in studying  close environment 
of the neutron stars. In this system a fast recycled $P_A= 22.7$ msec Pulsar A orbits
a $P_B=2.77$ sec Pulsar B on a 2.4 hour orbit inclined at $\sim 87.7^\circ$ (with 
uncertanity
of the order of unity)
to the line of sight. Pulsar A shows $27$ sec {\it frequency independent}
 eclipses at the moments
of superior conjunction \citep{kaspi}. In addition, Pulsar B shows orbital-phase dependent variations 
of intensity. It is also  very weak at the inferior  conjunction, consistent with
being  eclipsed as well (S. Ransom, private communication).  

\begin{figure}[ht]
\includegraphics[width=0.9\linewidth]{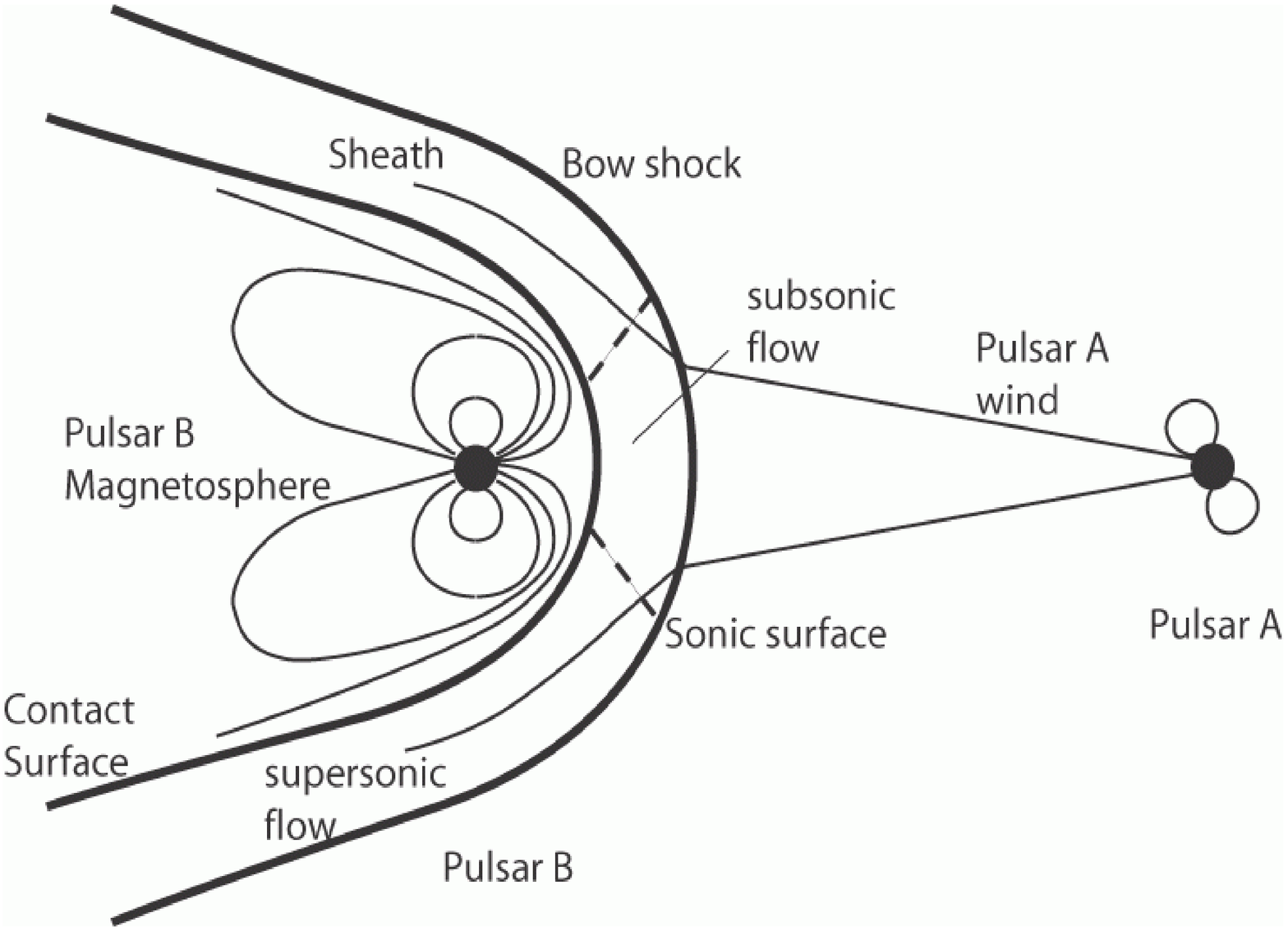}
\caption{ 
Cartoon of the binary pulsar system  PSR J0737-3039.
The wind of Pulsar A is shocked near Pulsar B and forms a sheath. 
Magnetosphere of Pulsar B is strongly distorted due to interaction with the wind.
The system resembles the Earth-Solar wind interaction. On the "day side" of  Pulsar B
magnetosphere (facing Pulsar A) the opening angle of the last closed field line
 may be much larger than in the case of isolated pulsar. In the sheath, the flow near
the apex point is subsonic, while at larger angles it becomes supersonic both due to 
changing conditions at oblique shocks and due to pressure acceleration along the sheath.
  The contact surface
is also  expected to be unstable to Kelvin-Helmholtz instabilities. }
\label{Binarypsr}
\end{figure}

In this paper we first  construct a geometrical model of  Pulsar A eclipses. 
Similarly to Arons \etal (private communication) we propose that the  interaction of
Pulsar A wind with   Pulsar B magnetosphere resembles Solar wind -- Earth magnetosphere
interaction (see Fig \ref{Binarypsr}). 
As a result, a bow shock forms in Pulsar A wind, separated from 
Pulsar B magnetosphere by a sheath of shocked material. 

Absorption of Pulsar A radio emission  may occur 
either in the shocked Pulsar A wind or in the magnetosphere of Pulsar B. 
If   absorption  occurs at cyclotron  resonance with
   non-relativistic particles present
on the  closed field lines, 
then it is expected to be strongly frequency dependent, which is not 
observed. Synchrotron
 absorption by relativistic particles on the open field lines
can give a frequency independent eclipse. Still, we disfavor this possibility,
  First, there are no indication of  modulation of the eclipse by the rotation
of  Pulsar B (though the present temporal  resolution may not be sufficient).
 Secondly, since the line of sight passes at the edge of magnetosphere
(see Fig.  \ref{pulsarB}), and since we see pulses from Pulsar B, it is hard to imagine
geometry in which near the superior  conjunction
 the line of light constantly  passes through open field lines, giving 
full eclipses.
Thirdly, the dimming of Pulsar  B near the
inferior conjunction is consistent with absorption being  due to the shocked Pulsar A wind.

An alternative possibility for the location of the 
absorbing material, which we favor, is that the absorption occurs in the 
shocked Pulsar A wind, enveloping  Pulsar B.
Though the size of the shocked sheath   is expected to 
vary by about $25\%$ due to Pulsar B rotation,
 it  may still have an approximately
constant column density and optical depth  regardless of its size.

 What determines the size
of the eclipsing region? It may be determined either by geometric factors,
so that the  beginning and the end  of the eclipse occur at the moment when the line of sight enters
the absorbing region, 
or by microphysical processes in the plasma so that  
 the eclipse  occurs at the moment when  optical depth  $\tau$ to scattering/absorption
becomes of the order of unity. We strongly prefer the
geometrical interpretation of  eclipses since  any relevant  microphysical process 
gives  a frequency dependent absorption coefficient so that the 
condition $\tau \sim 1$ would  occur at different points for different frequencies.

\begin{figure}[ht]
\includegraphics[width=0.9\linewidth]{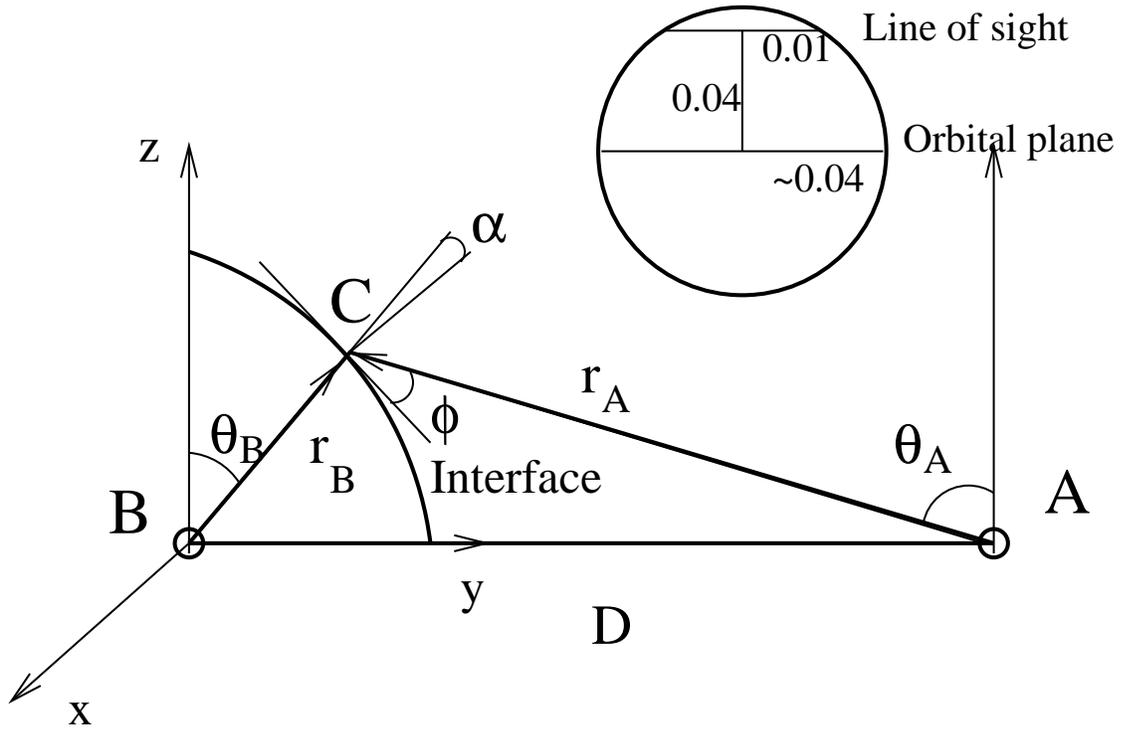}
\caption{
Geometry of the interaction. Pulsar A is located at distance $D$ from Pulsar B along the
$y$ axis. Magnetic moment of Pulsar B  may be oriented along any of the axis $x,y,z$. 
Pulsar A wind is  shocked  near Pulsar B. At point C on the interface surface,  defined  by
${\bf r}_B(\theta_B)$, the  surface normal makes an angle $\alpha$ with ${\bf r}_B$.
At the same point,  the radial flow from Pulsar A is inclined at angle $\phi$ to the surface.
The insert shows the eclipsing region (in radians) as viewed from Pulsar A.
}
\label{geom}
\end{figure}

The 
geometrical (as opposed to microphysical) interpretation  of the eclipse duration
seems, at first sight, inconsistent with the data:
a simple scaling of the size of Pulsar B magnetosphere would predict 
an eclipse which is much longer than observed.
 The light cylinder radius of  Pulsar B is $1.29 \times 10^{10}$
cm, so that its angular size seen from Pulsar A is $0.14 $ rad.
The line  of sight passes $2.3^\circ=0.04$ rad $=3.6 \times 10^9$ cm away from Pulsar B.
Eclipsing part, corresponding to a 27 sec eclipse duration, is $1.8 \times 10^9$ 
 cm \cite{kaspi},
\footnote{ 
 At this point magnetic field
of the B pulsar $B_B = 13$ G, so that the cyclotron frequency 
$\nu_B = 4.4\times 10^7$ Hz, which is too low to produce cyclotron absorption
if particles are non-relativistic.}
which subtends a total  angle of $0.02$ radians (see Fig. \ref{geom}). 
Thus, the size of the eclipsing region is $\sim 0.04$ rad $= 4 \times 10^9$ cm, which
is more than 3 times smaller than the light cylinder radius of  Pulsar B. 

 A possible resolution of this disagreement is that the pressure of 
Pulsar A wind compresses Pulsar B magnetosphere to  lateral sizes much smaller 
than the light cylinder. In this paper we first calculate 
 semi-analytically the form of the interface between  Pulsar A wind and 
Pulsar B magnetosphere. We find that if the magnetic field of  Pulsar B 
is calculated using the  vacuum dipole formula, the  lateral size of the  sheath
is still much larger than the one  inferred from eclipse duration. 
On the other hand, electromagnetic interaction of  Pulsar B with the sheath 
strongly modifies the structure of the magnetosphere, so that the vacuum 
dipole formula becomes inapplicable. The modification of the 
spin down torque leads to an estimate of   Pulsar B magnetic field which is typically 3-5 times smaller.
Still, the size of the sheath is much larger than inferred from eclipse duration. 
A possible resolution is that inclination angle is $\leq 86^\circ$, so that the line of sight 
cuts at the edge of the sheath. 

As  independent parameters of the system we chose the  spin down luminosities
$L_A = 5.8 \times 10^{33} $ erg/sec and $L_B = 1.6 \times 10^{30}$  erg/sec
and the separation of $D=9\times 10^{10}$ cm. These are the most reliable 
parameters, while  the
 quantities like the inferred  surface magnetic fields depend
on the details of the magnetospheric structure.

\section{Model of Pulsar A wind --  Pulsar B magnetosphere interaction}
\label{sec2}

Consider a point C on the interface between   Pulsar A wind  and
Pulsar B magnetosphere  defined by the radius-vector ${\bf r}_B(\theta_B)$ 
(see Fig.~ \ref{geom}). A normal to the interface makes an angle 
\be
\tan \alpha = \partial _{\theta_B} \ln r_B(\theta_B)
\ee
with the radius-vector ${\bf r}_B(\theta_B)$. Here 
 $\theta_B$ is the polar angle between $ {\bf r}_B$
 and the normal to the plane of the orbit.
At the same point, the radial flow from Pulsar A, 
\footnote{Pulsar A wind may be assumed radial since the wind velocity, which is of the order of the
speed of light, is much larger than the orbital velocity, $\sim 300 {\rm km s^{-1}}$ and 
much larger than the angular rotational velocity of the  wind, $v \sim c^2 /(D \Omega_A) \sim 0.001 c$,
where $\Omega_A= 2 \pi/P_A$ is the angular rotation frequency of Pulsar A.}
 leaving at the polar 
angle $\theta_A$, is inclined at angle $\phi$ to the interface.
Angles $\alpha, \phi, \theta_A$ and $\theta_B$
are related by
$\phi = \pi/2 + \alpha -( \theta_A+\theta_B)
$.  
From the triangle $ABC$ we also find
\be
\tan \theta_A + \tan \theta_B={D \over r_B \cos \theta_B}
\ee
and
\be
{r_A \over r_B} = { \cos \theta_B \over \cos \theta_A}
\ee

We assume that the position of the 
interface is determined by the pressure balance between the 
static pressure of Pulsar B magnetic field and the dynamic pressure of 
Pulsar A wind (this is often  called Newtonian approximation) :
\be
{L_A \sin ^2 \phi \over 4 \pi r_A^2 c} =
{\mu^2 \over 8 \pi r_B^6} f(\theta_B)
\label{1}
\ee
where $\mu$ is the magnetic moment of the B pulsar and $f(\theta_B)$
depends on the inclination of the magnetic moment. 
$ f(\theta_B) = 1,~ 1+ 3  \sin^2 \theta_B,~ 1+ 3  \cos^2 \theta_B$
for magnetic moments oriented along $x,y$ and $z$ correspondingly.

Next, we dimentionalize the problem by introducing
$r_B \rightarrow r D$, $\xi = 2  L_A D^4 /\mu^2$,  
\be
{f(\theta_B) \over r^6} 
= { \xi  \over 1+ ( \partial _{\theta_B} \ln r_B(\theta_B)) ^2}
  \left( { r + \cos \theta_B  \partial _{\theta_B} \ln r_B(\theta_B) - \sin \theta_B \over
 1 - 2 r \sin \theta_B + r^2} \right)^2,
\ee
which for a given 
$f(\theta_B)$ is  an ODE for $r(\theta_B)$ depending on $\xi$ as a parameter. 
Next, instead of $\xi$ we introduce a dimensionless (measured in terms of $D$)  stand-off distance $r_0$
\be
\xi = { ( 1 - r_0)^2 \over r_0^6} f(\pi/2)
\ee
and parametrize $r \rightarrow  r_0 g(\theta) $.
This gives
\be
{f(\theta_B) \over f(\pi/2)}=
{g^6 \over g^2 +g^{\prime 2}}
\left( { (1-r_0) (r_0 g^2 - g \sin \theta_B + g' \cos \theta_B ) \over 
1+ r_0^2 g^2 - 2  r_0 g \sin \theta_B } \right)^2 
\label{main}
\ee
For a given  stand-off distance $r_0$
 Eq. (\ref{main}) determines the form of the interface.
We integrate this equation numerically,  limiting integration
to the regions close  to  Pulsar B light cylinder 
(near the apex point $ g \sim  1+2 (\pi/2-\theta_B)^2/3$). 
Further down the stream the
 approximation of a static magnetosphere of Pulsar B breaks down
and the balance at the contact discontinuity is  determined by the 
pressure balance of two winds.

\section{Magnetic field of  Pulsar B}
\label{sec3}

In order to find the stand-off distance we need to estimate a magnetic field strength of 
 Pulsar B.
This is not straightforward since the conventionally used  vacuum dipole formula  is not applicable in this
case, as we argue below.

\subsection{Vacuum dipole spindown}

Conventionally,  magnetic field  of pulsars is determined by the vacuum dipole
 spin down formula
\be 
L=
  {2 \over 3} B_{NS}^2 R_{NS}^2 c \left( { R_{NS} \Omega \over c} \right)^4
\label{L}
\ee
Even in  case of isolated 
 pulsars this is only an approximation since a large contribution 
to the  torque comes from the  currents flowing on the open field lines 
of magnetosphere.( If the typical current density is the Goldreich-Julian
density times velocity of light, then the torque from the ${\bf j} \times {\bf B}$
force integrated over the open field lines gives the same estimate as  the 
vacuum dipole luminosity (\ref{L}).)

Equating (\ref{L}) to the spin down luminosity $L_B = \Omega \dot{\Omega} I $, where 
$I \sim 10^{45} {\rm g cm^2}$ is the moment of inertial of  neutron star,
gives  $B_B=1.5 \times  10^{12}$ G \citep{lyne}.
In this case  $\xi = 4 \times  10^7$,  $r_0 = 8.6 \times  10 ^{-3}$ for $f(\pi/2)=1$ (when the B pulsar
magnetic moment is oriented along $x$ or $z$ axis),  and
$r_0 =1.07 \times  10^{-2}$ for  $f(\pi/2)=4$  (when the B pulsar
magnetic moment is oriented along $y$ axis).
The corresponding stand-off distances are $r_m=4.8 \times 10^9$ cm and $r_m=6 \times 10^9$ cm.

For these parameters $r_0$ we integrate Eq. (\ref{main}). The results are given by upper
set of curves in Fig \ref{pulsarB}.  It is evident that the resulting size of the 
interaction sheath is much larger than the observed duration of the eclipse
(the expected duration of the eclipse is $\sim 10 $ times longer than observed). 

\subsection{Modification of the spindown of Pulsar B due to interaction with
Pulsar A wind}
\label{sec:main}

Applicability of the vacuum spin down formula  (\ref{L}) in the  case of  interacting system
PSR J0737-3039B is even more doubtful than in the  case of isolated pulsars. 
Since the magnetospheric radius is located deep inside the light cylinder, the 
structure of Pulsar B magnetosphere is strongly distorted if compared with the
vacuum case. The details of the structure are bound to be complicated. 

Qualitatively, there are three possibilities for the structure of Pulsar B magnetosphere
and corresponding spindown torque. (i) the interface may be a perfect conductor so that
no Pulsar B field lines penetrate it. (ii)  the interface may be   a perfect  resistor, 
so that all the Poynting flux  from Pulsar B reaching it gets dissipated. 
(iii)  the interface may be partially resistive so that large surface currents are 
generated, which together with the poloidal field of Pulsar B produce a torque
on the star.

Modification of the spindown torque in all these cases are very different. 
 We are mostly interested in cases (ii) and (iii).
For case (ii), assuming that the typical current on the 
open field lines is of the order of the Goldreich-Julian current and that the size of the
open field lines is determined by the magnetospheric radius $r_m$, 
the spin down  becomes
\be
L = \left( {2 \over 3}  \right)  B_{NS}^2 r_{NS}^2  c \left( { \Omega  r_{NS} \over c}\right)^2
\left( { r_{NS} \over r_m } \right)^2
\ee

\begin{figure}[ht]
\includegraphics[width=0.9\linewidth]{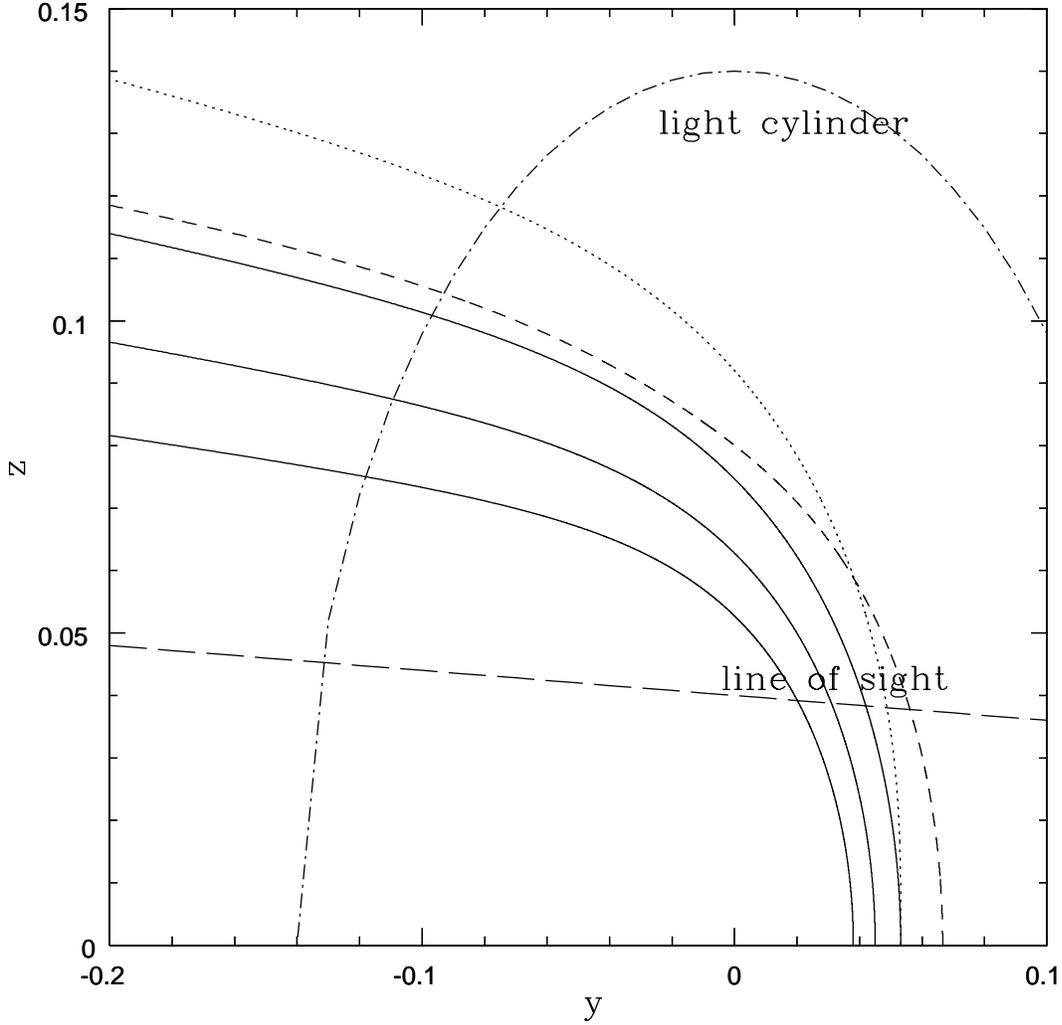}
\caption{  
Form of the contact surface in the $y-z$ plane for different orientations of  Pulsar B
magnetic dipole. Pulsar B is located at the origin, Pulsar A is located at
$y=1$.  Upper curves for magnetic field  of Pulsar B 
assuming standard spin down luminosity ($B_B=1.5 \times 10^{12}$ G).
Solid lines are  for $\mu=\mu_x$ (in the plane of orbit along the  Pulsar B orbital velocity),
dashed lines are for $\mu=\mu_y$ (along the line connecting the two pulsars)
and dotted lines for  $\mu=\mu_z$ ( perpendicular to the
orbital plane; observationally, large $\mu_z$ is excluded since we do see the pulses from
Pulsar B). Asymptotic opening angle is $\sim 0.11 -0.13$ rad.
Lower solid curves  for   magnetic field  of  Pulsar B with modified torques 
$B_B= 4.7 \times 10^{11}$ (Eq. \ref{Bn}) (asymptotic opening angle is $\sim 0.09$ rad)
and 
$B_B= 3\times 10^{11}$ G (Eq. \ref{Bn1}) (asymptotic opening angle is $\sim 0.07$ rad).
Long dashed line is the line of sight (inclination $87.7^\circ$).
Dash-dot line is the light cylinder radius (it looks non spherical due to
different axial scales).}
\label{pulsarB}
\end{figure}

Equating (\ref{L1}) to spin down luminosity $L_B$ and using the  force balance  
(\ref{1}), 
we find
\ba &&
B_{B} = { 3^{3/4} \over 2} { \sqrt{D} L_B^{3/4} \over L_A^{1/4} r_{NS}^3 \Omega^{3/2} } =
4.7 \times 10^{11} \, {\rm G}
\nn &&
 r_m =  { 3^{1/4} \over \sqrt{2}} \left( {L_B \over L_A} \right)^{1/4} 
\sqrt{ { c D \over \Omega}} = 4 \times 10^9\, {\rm cm}
\label{Bn}
\ea
Using parameters (\ref{Bn}) we integrate Eq. (\ref{main}) (the middle  solid   curve
on Fig. \ref{pulsarB}; in this case $r_0=0.045$ for $\mu =\mu_x$).
The size of the sheath is only slightly modified if compared with the
vacuum dipole 
and is  much larger than the one implied by eclipse
duration.

Finally,  the interface may be  partially resistive.  Relativistic boundary layers
are strongly unstable on scales 
of tens to hundreds of gyroradii \citep{su96,Liang}, so that
for ultra-relativistic electrons
 with $\gamma \sim 10^6$ the kinetic thickness of the interface
may become macroscopically large. 
As a result,  Pulsar B field may penetrate the interface, similar 
to what may happen in neutron star -- disk interaction \citep[\eg][]{gl79a,gl79b,w95}.
In this case the poloidal magnetic field of Pulsar B will  be twisted
by the pulsar rotation to produce toroidal magnetic field which may be as high as poloidal  magnetic field at the interface,  $B_\phi(r_m) \leq B_B(r_m)$.
(Since the magnetospheric radius is much smaller than the light cylinder radius, 
light travel effects
may be neglected.)
The resulting surface current on the interface together  with the
poloidal field will produce a ${\bf j} \times {\bf B}$ force and a torque on the pulsar.  
  The spin down then  becomes
\be
L =\left({1\over 2}  \right) 
\left( {2 \over 3}  \right)  B_{NS}^2 r_{NS}^2  c \left( { r_{NS} \over r_m } \right)^3
 \left( { \Omega  r_{NS}  \over c } \right)
\label{L1}
\ee
A factor $1/2$ has been introduced in front to account for the fact that
only the part of pulsar B magnetosphere facing  Pulsar A experiences the torque. 

Equating (\ref{L1}) to spin down luminosity $L_B$ and using the  force balance  
(\ref{1}), 
we find
\ba &&
B_{B} = { 3  \over \sqrt{ 2} }  { D L_B \sqrt{c} \over \sqrt{L_A} r_{ns}^3 \Omega}=
3 \times 10^{11}
~ {\rm G}
\nn &&
 r_m = \left({3 \over 2} \right)^{1/3}  \left({ L_B \over L_A} \right)^{1/3}
  \left({ c D^2 \over \Omega} \right)^{1/3} = 3.5 \times 10^9 {\rm cm}
\label{Bn1}
\ea
Magnetic field in this case is $\sim 5$  times smaller.

Using parameters (\ref{Bn1}) we integrate Eq. (\ref{main}) (the lower  solid   curve
on Fig. \ref{pulsarB}; in this case $r_0=0.038$ for $\mu =\mu_x$).
 
Both modified magnetic field  estimates (\ref{Bn}-\ref{Bn1})  predict 
 the size of pulsar B magnetosphere which  is 2-3 times larger than inferred by 
eclipse duration.  A possible resultion of the disagreement is that the 
orbital inclination angle is $ \leq 86^\circ$.

\subsection{Spin down due to propeller effect and Magnus force on Pulsar B}

Interaction of  Pulsar B with the wind of Pulsar A may also
produce propeller effect, whereby the material of  
Pulsar A wind is flung off by the rotation of Pulsar B. 
As a result, there will be extra torque on Pulsar B 
produced by the  Magnus force 
(a force due to a difference in pressures at the two sides of the
Pulsar B magnetosphere with  rotation velocity 
 aligned and counter-aligned with the radial wind of Pulsar A,
  c.f. a "dry leaf" kick in soccer).
Next we estimate the importance of the  propeller effect on the
spin down of 
Pulsar B. 

The torque of the Magnus force acting on Pulsar B can be estimated as  
\be
T= { L _A \Omega^2 r_m^5  \over 4  D^2 c^3 }
\ee
Equating this to the change of angular momentum we find the corresponding
magnetospheric radius
\be 
r_m = 2^{2/5} \left( { L_B \over L_A} \right)^{1/5} 
\left( {c^3 D^2  \over \Omega^3} \right)^{1/5} =
7.2 \times 10^9 {\rm cm}
\ee
This radius is larger than the one given by the dipole formula.
Thus, we conclude that the propeller effect is not important 
for the spin evolution of Pulsar B.
\footnote{The same Magnus force also produces  a torque on the orbital motion
of Pulsar B.
The corresponding effect on the orbital evolution is  too  small to be of any 
importance.}

\section{Eclipsing  mechanism: synchrotron absorption}

We propose that  the absorption of Pulsar A radio beam
occurs due to the synchrotron absorption in the shocked plasma.
We parametrize Pulsar A wind by a magnetization  parameter $\sigma$, the ratio of Poynting to particle
fluxes  \citep{kc84a}. The 
magnetic field in the sheath  (in the laboratory frame) is 
\be 
B = 3  \sqrt{ {\sigma \over 1+ \sigma} } \sqrt{ 2 L \over c D^2}
= 21 \sqrt{ {\sigma \over 1+ \sigma} } \,  {\rm G} = 1.2  \,  {\rm G}
\ee
where the last equality assumes $\sigma = 3 \times 10^{-3}$, a value inferred for Crab pulsar
\citep{kc84a}. The factor of 3 in front assumes a compression in strong relativistic 
fluid shock (see also Section  \ref{sec:obliq} for more details).
The 
 non-relativistic cyclotron frequency  inside the sheath is $\nu_B = 2.4 $ MHz.
In order to absorb at the observed frequencies
 $\nu_{obs} \sim 1$ GHz, the particles should be relativistic
with a Lorentz factor 
$\gamma \sim \sqrt{\nu_{obs} / \nu_B} =30$. {\it 
It is not obvious at all that such particles are present in the shocked flow.  }
Estimates of the bulk Lorentz
factor in case of the Crab pulsar give $\Gamma \sim 10^6$ \citep{kc84b}.
If a similarly high  Lorentz
factor is assumed for PSR J0737-3039A, then 
one expects that the lowest energy cut-off
in the shocked flow is similarly high, $\gamma \sim \Gamma$. 

On the other hand, radio emission of the Crab nebula  is attributed to  electrons with much lower
energies,  $\gamma \ll  \Gamma$. These  electrons  may be either 
  cooled remnants of the intense early injection \cite{atoyan99},
or, more likely, may  represent a completely different population of accelerated
particles. Thus, invoking the Crab pulsar as an example, it is feasible that
relativistic magnetized shocks do  produce a low energy population of electrons. 
Note, though, that since the magnetic fields in the bulk of the Crab nebular
and in the sheath of the interacting winds of  PSR J0737-3039 differ by some 4 orders
of magnitude, the corresponding Lorentz factors of synchrotron emitting/absorbing
particles differ by two orders of magnitude. Unfortunately, in the absence of understanding
of the acceleration mechanism of radio electrons we cannot judge whether this is possible.
(Alternatively, it is feasible that the low energy population in the sheath
 comes from Pulsar B due to a "leakage" through the interface.)

Below we assume that  a population of low energy electrons with a power
law distribution $n(\epsilon)= \kappa \epsilon^{-p}$  is indeed present 
in the sheath ($\epsilon$ is the energy of particles). 
The parameter $\kappa$ is related to the density of pairs and the low and
high energy cut-offs in the spectrum, $\kappa \sim  n m c^2 /(p-1) \gamma_{min}^{p-1} \sim n m c^2$
where we assumed that $\gamma_{min} \sim 1 $ and $p$ close to 2.

Cyclotron 
absorption coefficient  is then  \citep{lang74}
\be
\alpha_{\nu}^0 
 = 2 \times 10^{-2} \,
 g(p)\, (3.5 \times  10^9)^p\, \kappa \, B^{(p+2)/2}\,  \nu^{-(p+4)/2}\, {\rm cm^{-1}}
\label{mu}
\ee
where $ g(p)$ is a coefficient of the order of unity, $\nu$ is  observing frequency. 
We also assumed that radiation propagates orthogonally to the field lines. 

We normalize the pair density in the wind to the Goldreich-Julian density at Pulsar A 
light cylinder.  The particle flux for the  Pulsar A is then 
 $\dot{N}_A = \lambda  \sqrt{ 4 \pi L_A c /e^2} \sim  
\lambda \times  10^{32} s^{-1}$ and the  density 
at distance of  Pulsar B is $n_A= \dot{N}_A/(4 \pi c D^2) = 3\times 10^{-2} \lambda$ cc.
In order to estimate the thickness and the particle density in the sheath it is necessary to
know the details of the flow structure. As  a simple estimate we assume that the
column density through the sheath is of the order of the column density
of   Pulsar A at the distance of Pulsar B, $\sim n_A D$.
For particle index $p=2$ we find
\be
\tau_0 \sim  3 \times 10^{-6}   \lambda  \left({\nu \over 1GHz}\right)^{-3}
\label{tua0}
\ee
Thus, in order to produce optical depth $\geq 1 $ the  multiplicity factor 
should not be quite  large  $\lambda \geq 10^6$ (flatter spectra $p \leq 2$ give higher
optical depth also, for $p=1.5$ it increases by an order of magnitude).
 The required multiplicity is large, but not 
impossible (a value of multiplicity factor invoked for the Crab is $\lambda =10^6$,
\citep{kc84a}, see also \cite{arons79,musli03}).

We conclude that synchrotron absorption by weakly relativistic 
particles in the
shocked Pulsar A wind is a possible  
eclipsing  mechanism, but acknowledge that the required
wind multiplicity is at the upper possible end.

\section{Dynamics of the shocked Pulsar A wind}
\label{sec:obliq}

In the previous section we calculated semi-analytically the form of the
interface between Pulsar A wind and Pulsar B magnetosphere. 
In reality, the interface will consist of the bow shock and a contact
discontinuity (see Fig \ref{Binarypsr}).
\footnote{ In fact, since \Bf  of the  Pulsar A wind piles up
on the contact discontinuity, the two media are separated by a rotation
(\Alfven) discontinuity.} 
In this section we study the dynamics of the shocked pulsar wind 
in the sheath. 
We first derive the oblique jump conditions for 
relativistic MHD shocks. Relativistic oblique
 fluid shocks have been considered by \cite{koni80}; we are not aware of a work
which considered relativistic oblique MHD shocks. 
 We  apply the results to  
PSR J0737-3039, assuming that the form of the shock follows the form of the
interface calculated in Sections \ref{sec2} and  \ref{sec3}.

\subsection{Oblique fluid shocks}

\begin{figure}[ht]
\includegraphics[width=0.9\linewidth]{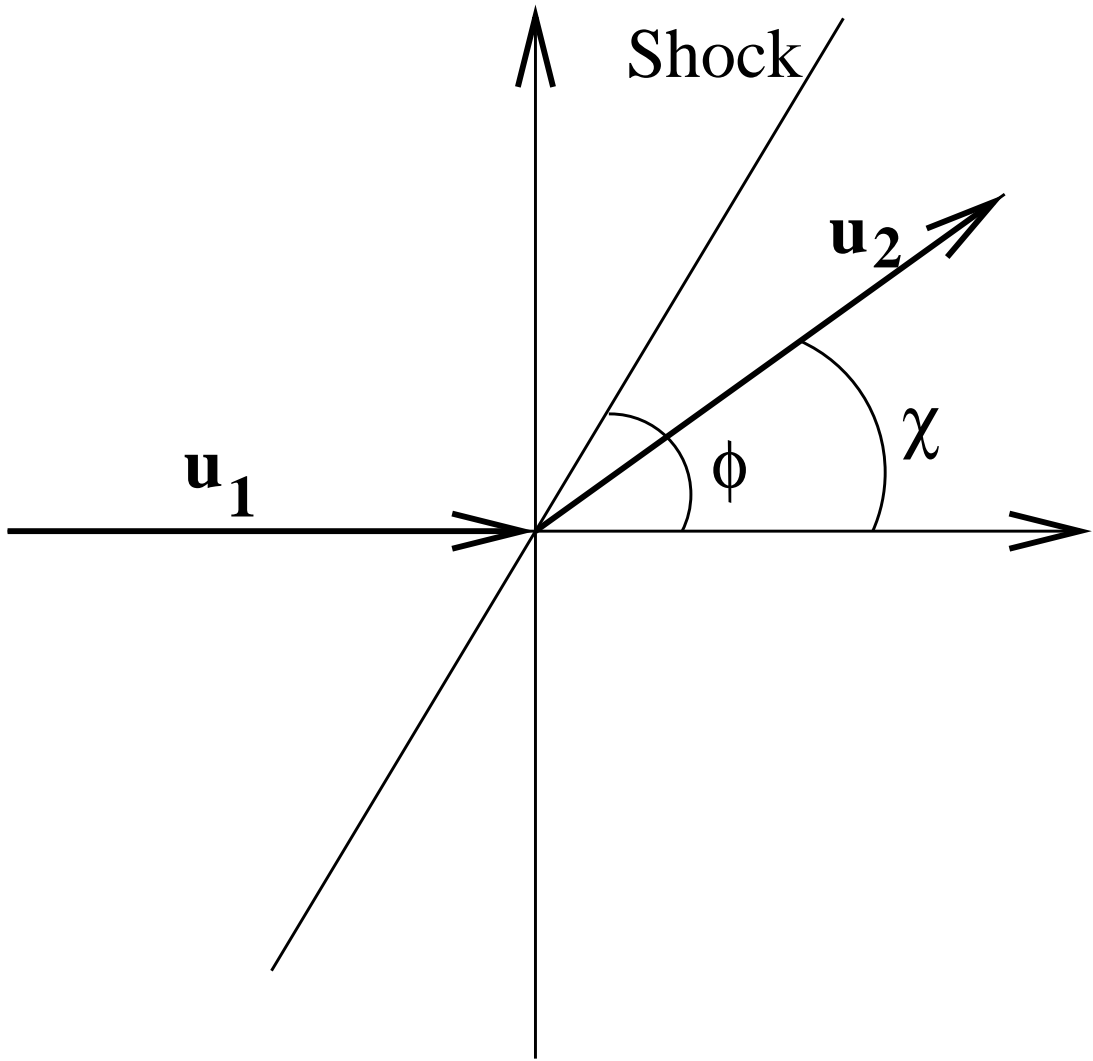}
\caption{Geometry of oblique shock flow.}
\label{geomRMHD}
\end{figure}

Let the stream lines make an angle $\phi$ with the shock 
and let the post-shock flow
make  an angle $\chi$ with the initial velocity (Fig. \ref{geomRMHD}).
Oblique shock conditions can be obtained from normal shocks and an additional condition
that the component along the shock velocity remains constant. 
\ba &&
n_1 u_1 \sin \phi = n_2 u_2 \sin(\chi - \phi)
\nn &&
w_1 u_1^2  \sin^2 \phi +p_1=w_2 u_2^2  \sin^2 (\chi - \phi) +p_2
\nn &&
w_1 \gamma_1 u_1  \sin \phi  =w_2 \gamma_2 u_2 \sin(\chi - \phi)
\nn &&
{ u_1^2 \cos ^2 \phi \over 1+ u_1^2 \sin^2 \phi}=
{u_2^2 \cos ^2 ( \phi - \chi)  \over 1+ u_2  \sin^2 ( \phi - \chi)}
\label{0}
\ea
where $w$ is enthalpy, $u$ is four-velocity, $n$ is density,
$p$- pressure and $ \gamma$ is Lorentz factor; velocity of light has been set to unity.
Subscripts denote
unshocked (1) and shocked (2) fluids.
Relations (\ref{0})
can be resolved \citep{LLIV} 
\ba &&
 u_1 \sin \phi  = \sqrt{ (e_2+p_1)(p_2-p_1) \over (e_1+p_1)(e_2-e_1 -(p_2-p_1)} 
\nn &&
 u_2 \sin(\chi - \phi) = \sqrt{ (e_1+p_2)(p_1-p_2) \over (e_2+p_2)(e_1-e_2 -(p_1-p_2)}
\nn &&
{n_1 \over n_2} ={u_2 \sin(\chi - \phi) \over u_1 \sin \phi } 
= \sqrt{  (e_1+p_2) (e_1+p_1) \over
(e_2+p_1) (e_2+p_2)} \equiv  \eta
\ea
where $w=e+p$ and $ \eta$ is the compression ratio, defined here  as
a ratio of  rest frame densities.

For adiabatic flow, using relations
\be
w= n + {\Gamma_a \over \Gamma_a -1} p = { u_s^2 +1 \over  u_s^2} p \Gamma_a
\ee
where $u_s$ is a sound four-velocity, and $\Gamma_a$ is adiabatic index.
 The compression ratio $\eta$
can be expressed as a function of $u_{s,1}$ and $u_{s,2}$.  The corresponding relations
are too bulky to be reproduced here. In the case of initially cold plasma
 ($p_1=0$, $w_1=e_1=n_1$, $u_{s,1}=0$)
we find
\ba  &&
u_1^2 = { (\Gamma_a -1) (\Gamma_a(1 - \eta) - (1+\eta) )
\over \Gamma_a ( 2- \Gamma_a(1 - \eta)) \eta^2 \sin^2 \phi}
\nn &&
u_2 =  \eta u_1 { \sin \phi \over  \sin (\chi - \phi)}
\nn &&
u_{s,2} ^2 =  { (\Gamma_a -1) (\Gamma_a(1 - \eta) - (1+\eta) )
\over  \Gamma_a (3 -  \eta) -  \Gamma_a^2 ( 1 - \eta) -2 - \eta }
\label{2}
\ea
In the non-relativistic limit $u \ll 1 $ these relations give
$\eta = (\Gamma_a-1)/(\Gamma_a+1) + { {\cal{O}} (u^2)}$,
while in the strongly relativistic limit $u_1 \gg 1$, $\eta \ll 1$,
\ba &&
\eta = { \Gamma_a -1 \over
u_1 \sin \phi \sqrt{ \Gamma_a ( 2 - \Gamma_a)}}  = { 1\over  2 \sqrt{2} u_1 \sin \phi }
\nn &&
u_2 =  { \Gamma_a -1 \over
\sqrt{ \Gamma_a ( 2 - \Gamma_a)} } = { 1\over  2 \sqrt{2}} 
{ \sin \phi \over  \sin (\chi - \phi)}
\nn &&
u_{2,s} = \sqrt{  \Gamma_a -1 \over 2 - \Gamma_a } = {1 \over \sqrt{2}}
\label{3}
\ea
where the second equalities assume $\Gamma_a =4/3$ (this is also assumed 
for all numerical estimates below). Thus, for cold flow
$0< \eta < (\Gamma_a -1)/( \Gamma_a+1)$.

As independent parameters of the problem we chose the initial
four-velocity $u_1$ and the compression ratio $\eta$. For a particular
case of cold plasma  $u_1$
and $\eta$ are related by  Eq. (\ref{2}).

Eliminating $u_2$ from Eqns.
 (\ref{0}), we can determine $\chi ( \phi, u_1)$:
\be
\tan^2 ( \phi - \chi) =  \eta ^2  {1+ u_1^2 \sin^2 \phi \over
1 +  \eta ^2 u_1^2 \sin^2 \phi} \tan^2   \phi
\ee
which, using Eq. (\ref{2}) gives
\be
\tan\chi = { \Gamma_a (1 -   \eta)  \tan  \phi \over
1 +  (1- \Gamma_a (1 -   \eta)) \tan^2  \phi}
\ee
For strong shocks, $\eta \rightarrow 0$,
\be
\tan\chi = { \Gamma_a   \tan  \phi \over
 1 -(\Gamma_a-1) \tan^2  \phi}
\ee
For $\Gamma_a =4/3$, the maximum deflection angle $\chi_{max} = \pi/6$
is reached at $\phi_{max} =  \pi/3$. 

We can also determine the post-shock Mach number:
\be
M_2^2 = {u_2^2 \over u_{2,s}^2} =
{u_1^2 \over u_{2,s}^2} { \cos^2 \phi + (1+ u_1^2)  \eta^2 \sin ^2 \phi
\over 1+ u_1^2 \sin^2 \phi } 
\approx { 2 \cos^2 \phi + 1/4 \over \sin^2 \phi}
\label{M2}
\ee
For initially cold flow, the compression ratio $\eta$ can be eliminated
using relations (\ref{2}); then Eq. (\ref{M2}) 
can be solved for the angle $\phi_M$ at which the post shock flow is sonic.
We find $\phi_M=\phi_{max} =  \pi/3$.

\subsection{Oblique MHD shocks}

Next we find jump conditions for strong,  ultra-relativistic fast magnetosonic shocks,
assuming
that the magnetic field lies in the plane of the shock.
(When the field is in the plane of the shock, the relevant
MHD expressions  can  be obtainable by generalizing the hydrodynamical 
 relations
by substituting for the pressure $p$  and  internal  energy density 
$\epsilon$: $p \rightarrow p+ b^2/8\pi $, $\epsilon\rightarrow \epsilon +
b^2/8\pi $.) The shock jump conditions for relativistic MHD shocks  are \cite{LLIV}
\ba &&
b_1 u_1 \sin \phi = b_2 u_2  \sin (\chi - \phi)
\nn &&
\left( w_1 + b_1^2 \right)  u_1 ^2  \sin^2  \phi +p_1 +{b_1^2 \over 2} =
\left( w_2 + b_2^2 \right)  u_2^2  \sin^2  (\chi - \phi)  +p_2 +{b_2^2 \over 2}
\nn &&
\left( w_1 + b_1^2 \right)  u_1 \sin \phi \gamma_1=
\left( w_2 + b_2^2 \right)  u_2  \sin (\chi - \phi) \gamma_2
\label{MHD}
\ea
(continuity equation and equation for velocity along the shock remain the same).
In Eq. (\ref{MHD})
 $ b $ is a rest frame magnetic field times $\sqrt{4 \pi}$.
Equations (\ref{MHD}) can be resolved
\ba &&
u_1 =  \sqrt{ \left( (b_1^2 - b_2^2)/2+p1-p_2  \right) \left( (b_1^2+  b_2^2)/2+p1-p_2+w_2 \right)
 \over 2 (b_1^2 + w1)(2(p_2-p_1) + w2-w1) } {1 \over \sin  \phi }
\nn &&
u_2 = \eta {  \sin \phi \over  \sin (\phi -\chi)} u_1
\nn &&
\eta=
{ n_1 \over n_2} =
{b_1 \over b_2} ={u_2 \sin (\phi -\chi)  \over u_1  \sin \phi} 
=\sqrt{\left( w_1 + b_1^2 \right) \left(w_1-p_1+p_2+ (b_1^2 + b_2^2 )/2 \right) 
\over 
\left( w_2 + b_2^2 \right) \left(w_2-p_2+p_1 + (b_1^2 + b_2^2 )/2 \right) }
\label{eta1}
\ea
To characterize the magnetization of the flow we introduce parameter
(see \cite{kc84a})
\be
\sigma= {b_1^2 \over w_1}
\ee
which is the ratio of the rest frame magnetic and plasma energy densities
(it is also equal to the ratio of Poynting to particle fluxes). 

Similarly to the fluid case, the compression ratio  (\ref{eta1})
can be written in terms of preshock and post-shock fast magnetosonic four-velocities
\be
u_s^2 = { p \Gamma_a + w \sigma \over w - p \Gamma_a }
\ee
If the preshock plasma is cold and 
 in the limit of strong shocks
 $\eta \rightarrow 0 $ (it is also necessary that $u_1^2 \gg \sqrt{\sigma}$),
 and assuming 
$\Gamma_a =4/3$,
we find
\ba && 
u_1= {1 \over \eta}
  \sqrt{ { 2  + \Gamma_a ( \sigma-2) - 4  \sigma - \sqrt{ \Gamma_a^2 (2+ \sigma)^2 
+ 4(1+ 2 \sigma)^2 -  4 \Gamma_a ( 2+ 3 \sigma + 2 \sigma^2) }  \over 
8 (2 - \Gamma_a) \left( \sqrt{ \Gamma_a^2 (2+ \sigma)^2
+ 4(1+ 2 \sigma)^2 -  4 \Gamma_a ( 2+ 3 \sigma + 2 \sigma^2) } + 2 + 
(4 - \Gamma_a) \sigma \right)  }}
=
\nn &&
 { 1 + 4 \sigma +\sqrt{1+ 16 \sigma( \sigma +1)} 
\over  2 \sqrt{2} \eta \sqrt{3+  4 \sigma +\sqrt{1+ 16 \sigma( \sigma +1)} }}
\equiv {f(\sigma) \over \sin \phi  \eta}
\nn &&
u_{2,s} =  \sqrt{\Gamma_a -1 +\sigma  \over 2- \Gamma_a } = \sqrt{ 1+ 3 \sigma \over 2}
\label{sigma}
\ea
(\cf \cite{kc84a}). 
Equation (\ref{sigma}) defines the function $f(\sigma) $. 

Equations describing oblique relativistic MHD shocks remain the same as in the fluid
case, with dependence $u_1(\eta)$ given by 
(\ref{sigma}). 
The deflection angle becomes
\be
\tan \chi = { ( f(\sigma) + \sqrt{1+ f(\sigma)^2}  ) \tan \phi 
\over \sqrt{1+ f(\sigma)^2} - f(\sigma) \tan ^2 \phi}
\ee
The maximum deflection angle is reached
\be
\cos \phi_{max} =  \sqrt{ f(\sigma)(  \sqrt{1+ f(\sigma)^2} - f(\sigma)) }
\label{phimax}
\ee
As $\sigma \rightarrow \infty$, 
 $\phi_{max}  \rightarrow \pi/4$, $\chi_{max}  \rightarrow 1/(4 \sigma)$.

\begin{figure}[ht]
\includegraphics[width=0.9\linewidth]{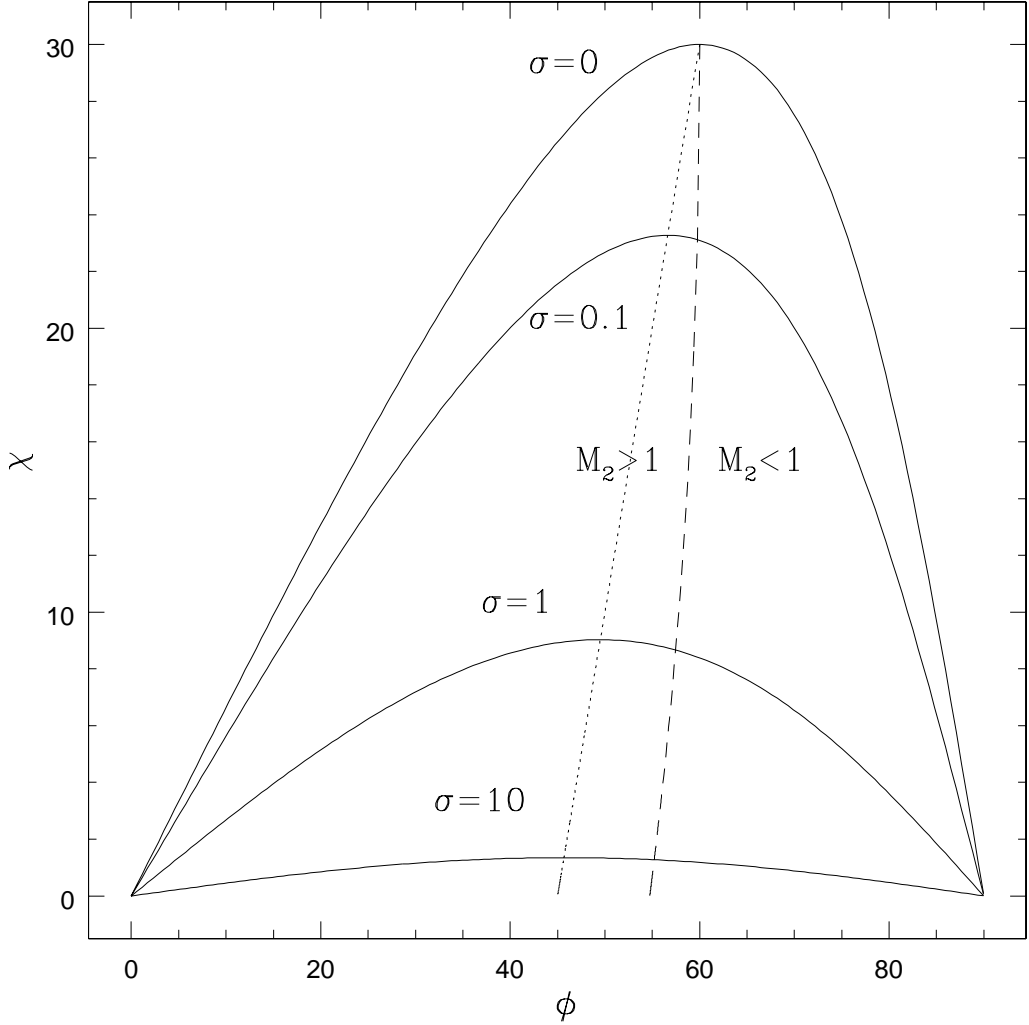}
\caption{Dependence of the deflection angle $\chi$ on the angle of attack
$\phi$ for ultra-relativistic MHD shocks $u_1 \gg 1$ and different magnetization
parameters $\sigma$. Dotted lines gives the maximum deflection angle 
(Eq. (\ref{phimax})), dashed line demarcates the subsonic shocked flow (to the right)
and supersonic flow (Eq. (\ref{phi1})).  For fluid shocks 
the maximum deflection  angle $\chi_{max}=30^\circ$ is reached
at $\phi_{max}= 60 ^\circ$. This is also the point when   the shocked  flow becomes
 subsonic.
}
\label{chimhd}
\end{figure}

Similarly to the hydrodynamic case, 
using Eq. (\ref{M2}), we can find the angle  $\phi_1$ at which the post shocked flow
remains sonic. In the ultra-relativistic limit we find
\be
\sin \phi_M = \sqrt{ (2 - \Gamma_a) (1+ f(\sigma)^2) \over 1+ \sigma}
\label{phi1}
\ee
(see Fig. \ref{chimhd}).
As $\sigma \rightarrow \infty$, $\phi_M  \rightarrow  \arcsin \sqrt{2 \over3} = 
54.73^\circ$, $\chi_M \rightarrow 1/( 3 \sqrt{2} \sigma)$. 

\begin{figure}[ht]
\includegraphics[width=0.45\linewidth]{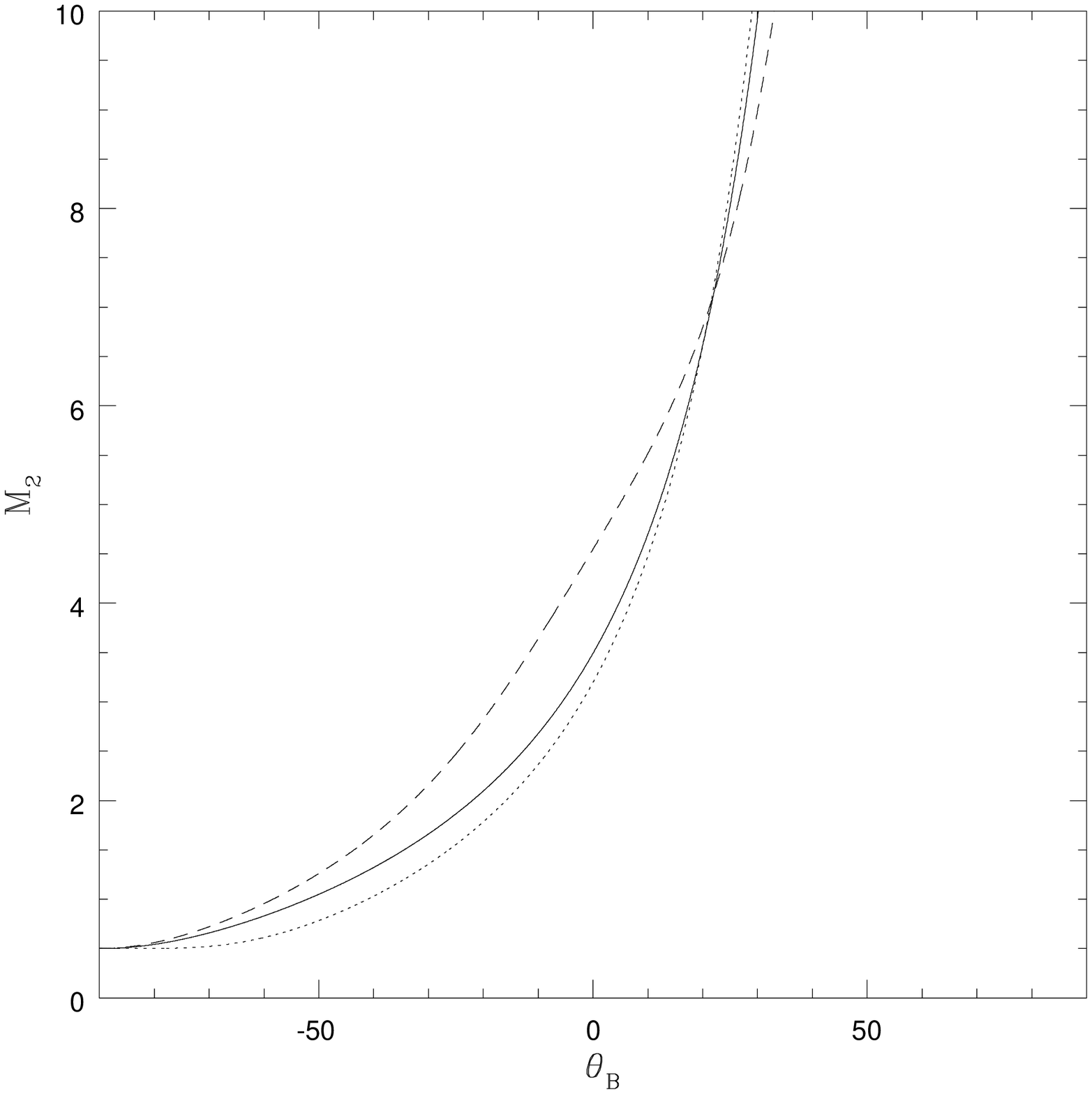}
\includegraphics[width=0.45\linewidth]{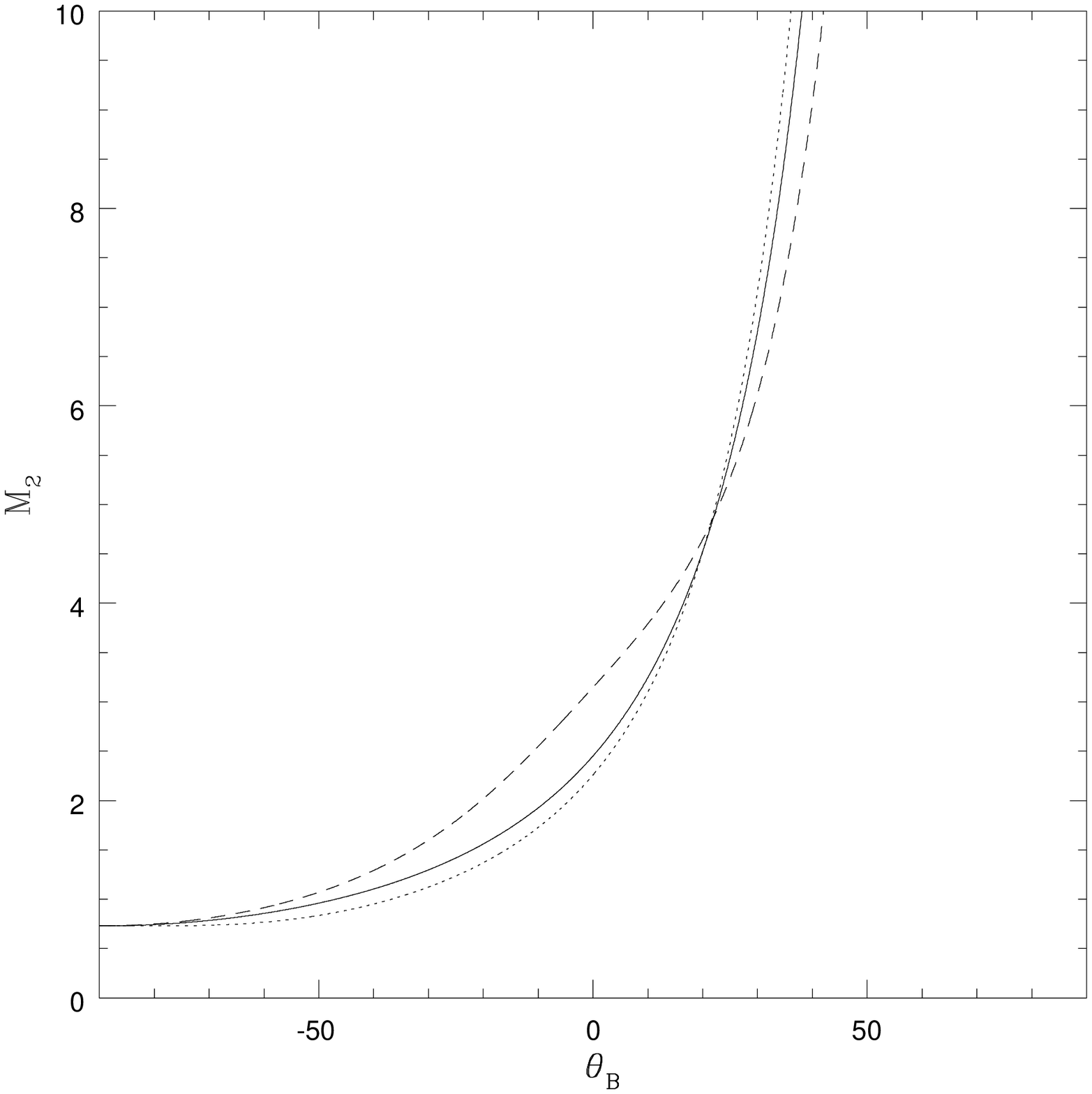}
\caption{
Post-shock Mach number  as a function of  polar angle. (a) hydrodynamic flow
$\sigma=0$ (post-shock four-velocity  is 
$ M_2/\sqrt{2}$), (b) magnetized wind $\sigma=1$
 (post-shock four-velocity
$ \sqrt{(1+3 \sigma)/2}  M_2$).  Labeling of the curves is the same
as in Fig. (\ref{pulsarB}).
} 
\label{M2b}
\end{figure}

Using the  relations derived in the previous section we can calculate the post shock velocities (Fig. \ref{M2b}).
At larger angles the post-shock flow becomes relativistic, $\Gamma_s \geq 1$
($\Gamma_s$ is the post-shock Lorentz factor).

\section{Pulsar B emission}

Another puzzling property of  PSR J0737-3039 is the variations of Pulsar B flux
 depending on the orbital phase \citep{lyne}.
In addition to two sections of the orbit where it is very bright, so that
single pulses can be seen, the flux density of the pulsar B
emission is at a minimum near, but centered slightly before,
the inferior conjunction. It is not clear if the flux goes to zero
(Ransom, private communication).
One possibility is that Pulsar A  wind leaks through the interface and affects
the microphysical process responsible for the generation of radio emission by Pulsar B,
somewhat similar to the  so-called flux transfer events at the day side
of Earth magnetosphere \citep{fahr86}.
 Such process is prohibited 
in ideal MHD and should occur through resistive effects (\eg tearing mode).
 The fact that the magnetized boundary becomes
``leaky'' and both plasma and magnetic field are transported across it,
 has been amply demonstrated through
decades of space experiments \citep[\eg][]{cowley82}. The transport occurs
 either due to microscopic resistive  instabilities
of the surface current \citep{gkz86}
or due to dynamic (\eg RT)  instabilities.

Can Pulsar A affect  Pulsar B electrodynamics? 
We have previously estimated the density of Pulsar A wind at  Pulsar B to be
$n_A \sim 
 3\times 10^{-2} \lambda$ cc.
For Pulsar B, the particle flux is  $\dot{N} = \lambda \sqrt{ 4 \pi c L_B / e^2}=
 10^{30} \lambda s^{-1}$ 
(surface Goldreich-Julian density  $n_{GJ, surf,B}=  2.2 \times 10^9 cm^{-3}$), so that the
density at the interface $r_m \sim 2 \times 10^9$ cm is $n_B\sim 0.27  \lambda$ cc.
Thus, if the multiplicity factors are equal for both pulsars, the 
particle
density of  Pulsar B flow on the open field lines is two orders of magnitude larger than
 the density of Pulsar A wind. Overall,  Pulsar A wind can make only 
a small contribution to the particle  density  in  Pulsar B magnetosphere.
On the other hand, 
the assumption of equal  multiplicity factors may not hold in the gaps of Pulsar B - regions
of low density, where $\lambda_B \sim 1$.  If Pulsar A plasma can get onto field
lines passing through  Pulsar B gaps, and if $\lambda_A \geq 100$, this  can strongly
affect particle acceleration in   Pulsar B gaps and radio emission generation. 

\begin{figure}[ht]
\includegraphics[width=0.35\linewidth]{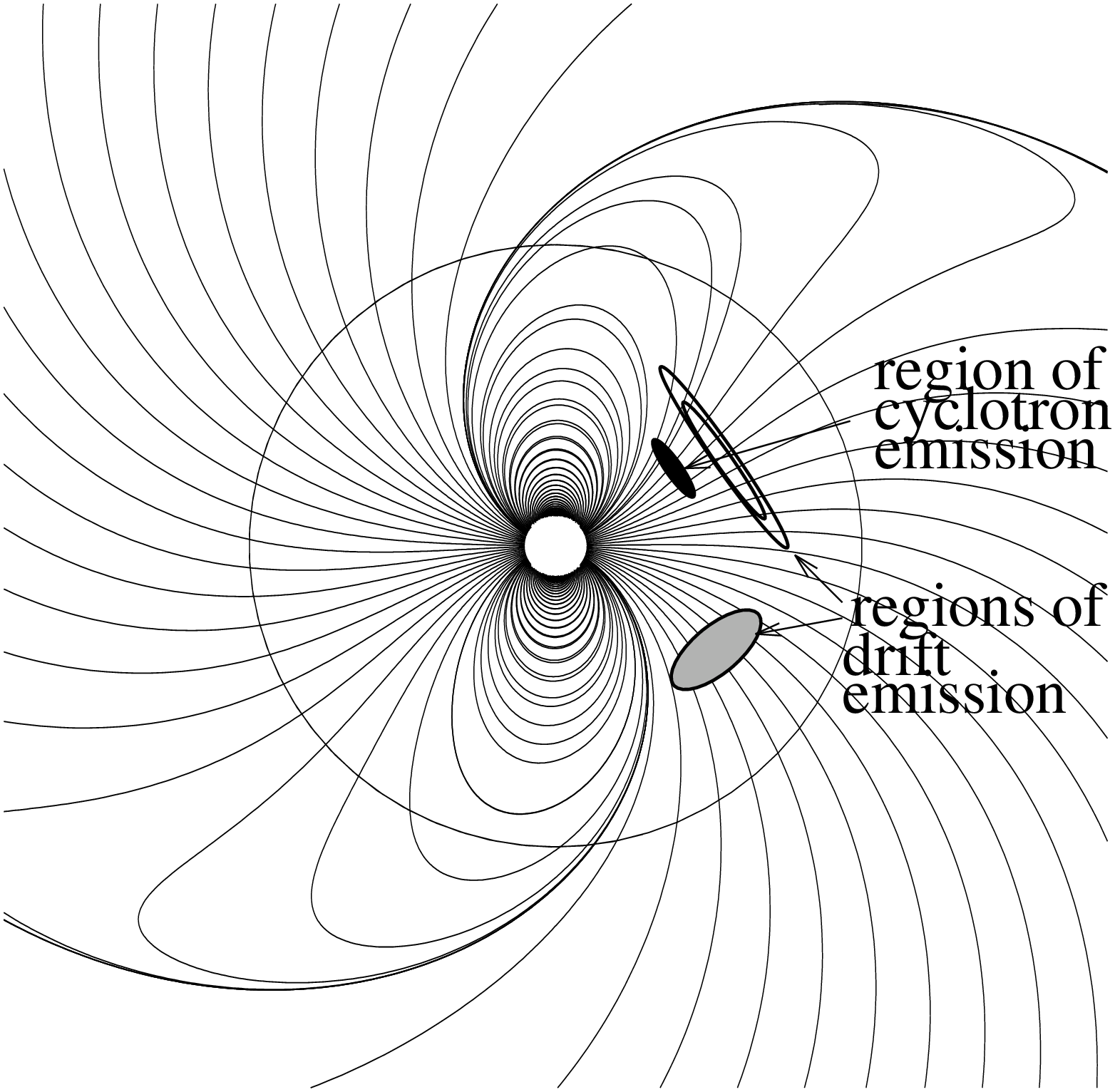}
\includegraphics[width=0.55\linewidth]{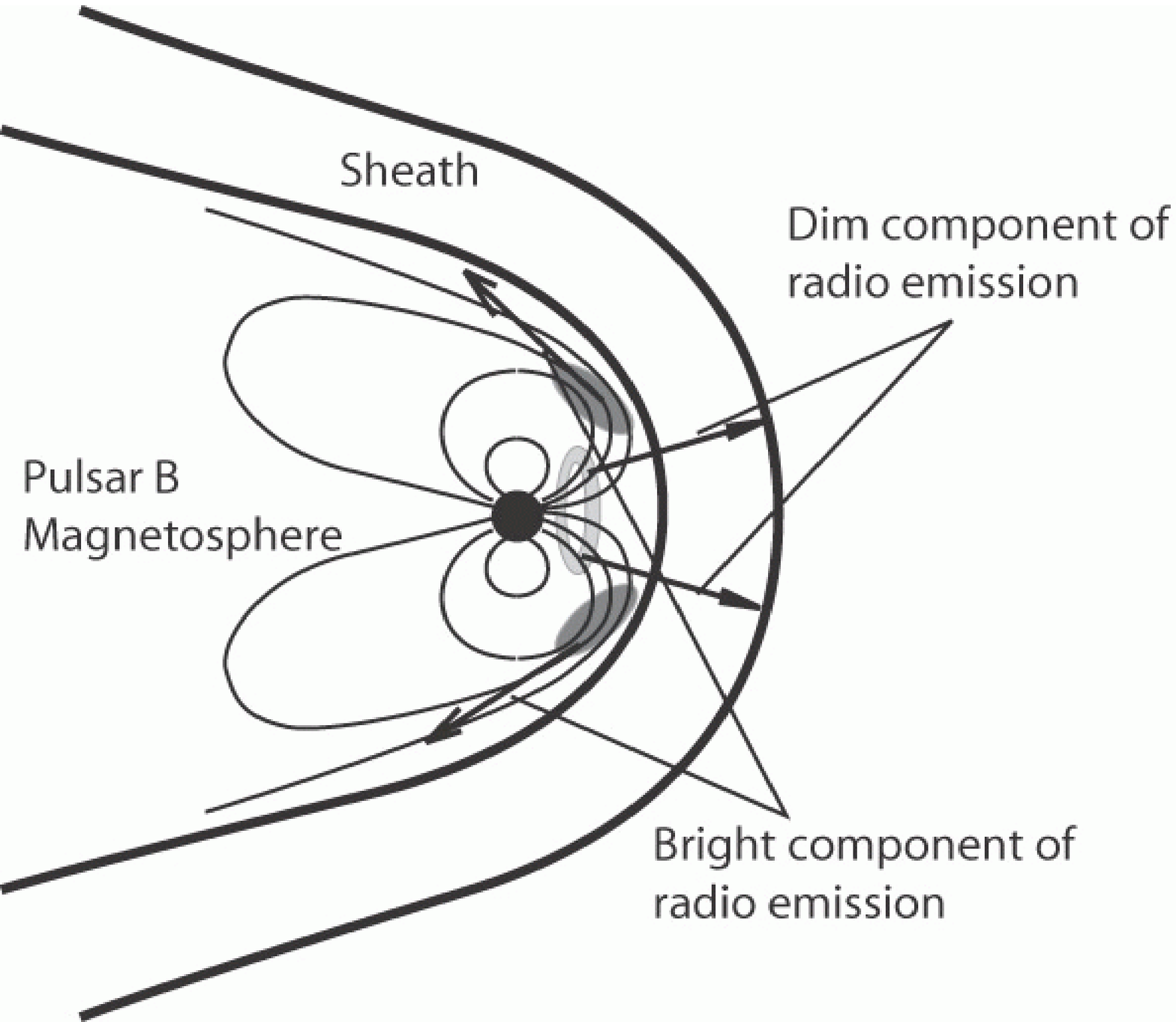}
\caption{(a) Fig. 5 from \cite{lbm} showing the location of radio emission generation regions
in isolated pulsars (field structure is that of a vacuum rotating dipole). 
 ''The location of the Cherenkov-drift emission depends sensitively on the
curvature of magnetic field line. Two possible locations
of the Cherenkov-drift emission are shown: ringlike near the
magnetic axis and in the region of swept field lines.''  
(b) Possible modification due to interaction with Pulsar A wind. 
The dim component of emission is not affected by the Pulsar A wind and is produced at all phases
of Pulsar B rotation, 
 while the bright component
is produced only at  the side of Pulsar B magnetosphere facing Pulsar A.
  }
\label{Bemis}
\end{figure}

Finally, we suggest a possible  explanation for  sudden increases in brightness of  Pulsar B
at two particular sections in the orbit.
The {\it  dynamical compression  of Pulsar A wind may change the condition
for radio emission generation if generation takes places at large distances from the pulsar,
as has been suggested
by \cite{lbm}} (see also \cite{kmm91}
 For  example,  the growth rate of 
 Cherenkov-drift  instability 
 \citep{lmb}   depends sensitively on the radius of curvature of  magnetic   field lines. 
Due to strong pressure from Pulsar A wind, the open field lines of Pulsar B are much stronger curved, 
than in isolated pulsars, producing larger growth rates (growth rates for
 Cherenkov-drift instability are marginal in isolated pulsars, \citep{lmb}).
Cherenkov-drift instability develops at  a limited  range of radii and produces emission which
is beamed along the local magnetic field. As \cite{lbm} have argued, 
emission is produced at two locations: 
in a ring-like region near the magnetic axis and in the region of swept back magnetic field lines.
In the latter case the  radio emission is produced at  
  large angles with respect  to the magnetic axis (see Fig.  \ref{Bemis}). 
We  propose  that the  transient brightening of Pulsar B occurs  when the line of sight
runs parallel to the magnetic field lines in the   emission generation region located  on the bend-back
field lines, close to the
edge of Pulsar B magnetosphere.

The immediate implication of the model is that  the pulse profile of
Pulsar B should change with the orbital phase, as the new emission region  comes into line
of sight at   particular parts of the orbit.
 Preliminary data indicate that this is indeed the case (S. Ransom, private communication).

An important prediction of the Cherenkov-drift instability, which is in stark contrast to the
the  bunching theory of radio emission, is that the  {\it emitted waves
 are polarized perpendicular
to the plane of the curved magnetic field line.} Thus at the swept-back field lines polarization
is along the axis of rotation. This may be
used as a test to distinguish between the two theories. Assuming that Pulsar B is an orthogonal
rotator with the rotational axis along the normal to the orbital plane,
the   Cherenkov-drift emission   will be linearly polarized along the normal to the orbit as well.

\section{Discussion}

In this paper we 
considered magnetohydrodynamic interaction between relativistic pulsar wind and
static magnetosphere in  double pulsar system
PSR J0737-3039. Our main conclusion are:
\begin{itemize}
\item   Electromagnetic torque on  Pulsar B is  increased due to the
 interaction with
Pulsar A wind. Depending on the nature of interaction,
the magnetic field of Pulsar B can be as low as $3 \times 10^{11} $ G.
 Still, the geometrical  model for the
form of the interface   predicts the size of the interface which is much larger than inferred from 
the 
 duration of Pulsar A eclipses, unless the orbital  inclination angle is $\leq 86^\circ$.
\item A likely 
 cause of eclipses is  synchrotron absorption in the dense shocked Pulsar A wind
by low energy relativistic electrons.  The density of Pulsar A wind should be at the 
upper allowed limit. 
The presence of such electrons cannot be justified
from first principles. 
\end{itemize}

The major uncertainty of the model is the source of low energy, $\gamma \sim 30$,
 electrons in the sheath. 
Pulsar A wind cannot be so slow: otherwise induced Compton scattering in the wind
will make it unobservable \citep{sk92}. One  possibility is  that such low energy
electrons are accelerate not at the shock, but at the rotational discontinuity
during development of dynamic and/or resistive instabilities. Alternatively, 
mixing of Pulsar A and Pulsar B plasmas, initiated by such instabilities, may populate the sheath with 
weakly relativistic electrons accelerated in Pulsar B gaps. 
If  mixing is efficient, particle number density in the sheath may be dominated
by  Pulsar B plasmas.  Yet another possibility is that
absorption happens on the strongly bend-back open field lines of Pulsar B which asymptotically
take the form of the sheath (it would be hard to distinguish this possibility from absorption
in the sheath itself).

Our calculations are consistent with the possibility that Pulsar A wind
experiences  a strong shock near Pulsar B. This limits the magnetization 
parameter to $\sigma \leq \Gamma^2$ (this is a condition that the preshock Lorentz
factor $\Gamma$ is larger than the \Alfven speed in the wind). 
Thus,  our results  {do not necessarily imply
that  Pulsar A wind is kinetically dominated near Pulsar B}. 
(The energy required to create a population of low energy electrons
is a $1/\sigma$ fraction of the total wind energy; this is also
a fraction of energy dissipated in a strongly magnetized shock with $\sigma \gg 1$,
\citep{kc84a}.)

A possible measurement of $\sigma$ may come from high energy observations of the system.
Shocks in 
kinetically dominated winds are expected to be strongly dissipative so that
a large fraction of the incoming energy flux may be radiated in X-rays. 
The expected luminosity is $L_X \sim L_A \Delta \Omega\sim 3 \times 10^{30}$ erg/sec,
where $\Delta \Omega \sim 5 \times 10^{-3}$ is the solid angle of  the shocked
region seen from Pulsar A.  A recent detection of a weak X-ray emission
\citep{mclaugh04} at a level $2 \times 10^{30}$ erg/sec is consistent with this scenario. 
(A confirmation of this result is needed since the count rate was too low and 
the X-ray emission is also consistent with Pulsar A magnetospheric emission.)

A better understanding of the system should come from the full relativistic MHD modeling 
of the system. The  relations for oblique relativistic MHD shocks  derived in this paper 
can be used  as guidelines and a  check to  such 
simulations. Qualitatively, the flow in  the sheath is expected to become supersonic both
due to the changing shock conditions and due to pressure acceleration away from the
apex point, so that the flow will form a de Laval nozzle. 
In addition, centrifugal forces acting on a flow moving along the
curved trajectory (so called Busemann correction) need to be taken into account. 
Another important property of the shocked flow is that even for small
magnetization of Pulsar A wind $\sigma \ll 1$ {\it magnetic field plays an important
role near the  contact discontinuity} and cannot be neglected.  Thus, any credible
simulation of the interaction {\it  must } use full relativistic MHD formalism.

\begin{acknowledgements}
I would like to thank Chris Thompson for his interest in this work and numerous 
discussions and Jon Arons for his critical and insightful comments. 
I also would like to thank 
  Roger Blandford, Arieh Konigl, Yuri Levin, Scott Ransom and Anatoly Spitkovsky.
I am also grateful to Alissa Nedossekina  for comments on the manuscript.
This research has been supported by NSERC grant  RGPIN 238487-01.
\end{acknowledgements}

{}

\newpage

\begin{figure}[ht]
\includegraphics[width=0.9\linewidth]{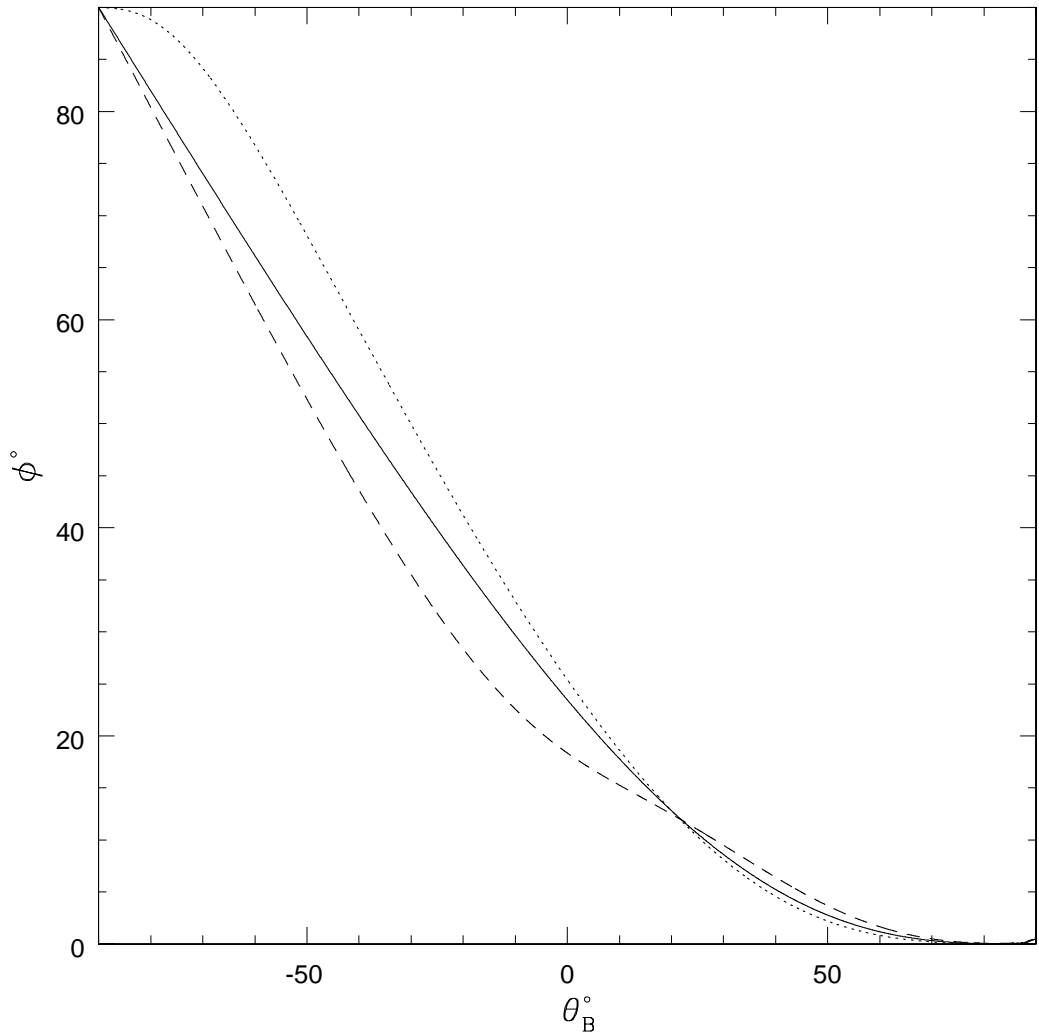}
\caption{
Angle of attack $\phi$ as a function of  polar angle. Negative $\theta_B$ refer
to parts of the interface facing  Pulsar A. Labeling of the curves is the same
as in Fig. (\ref{pulsarB}).}
\label{phi}
\end{figure}

\begin{figure}[ht]
\includegraphics[width=0.9\linewidth]{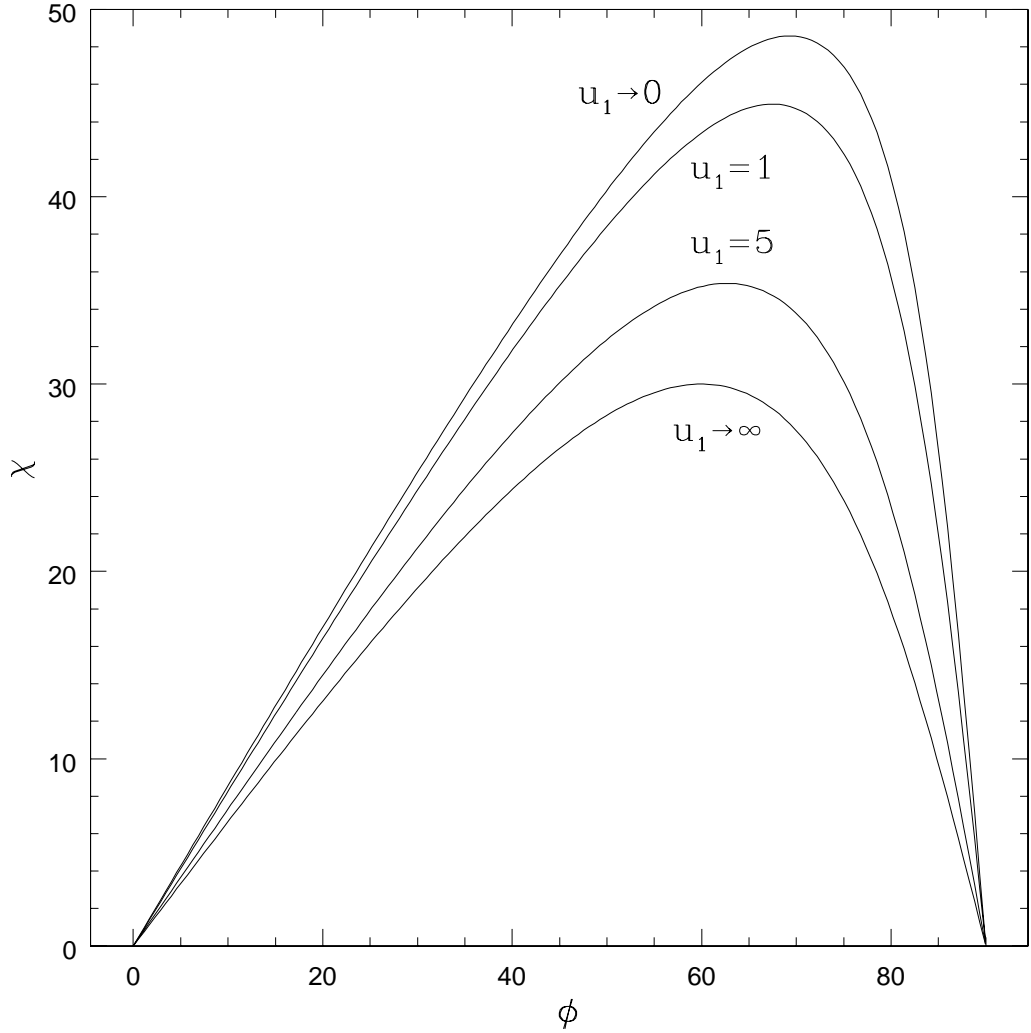}
\caption{Dependence of the deflection angle $\chi$ on the angle of attack
$\phi$ for initially cold ($p_1=0$) fluid for different $u_1$  and
 $\Gamma=4/3$. 
$u_1 \rightarrow 0$ corresponds to non-relativistic case.}
\label{chiu}
\end{figure}

\end{document}